\documentclass{amsart}
\usepackage{amssymb}

\swapnumbers
\theoremstyle{plain}
\newtheorem{thm}{Theorem}[section]

\newtheorem{prop}[thm]{Proposition}
\newtheorem{cor}[thm]{Corollary}

\newtheorem*{OpQu*}{Open question}
\theoremstyle{definition}
\newtheorem{defn}[thm]{Definition}
\theoremstyle{remark}
\newtheorem{note}[thm]{Note}
\newtheorem*{note*}{Note}

\newtheorem*{exmp*}{Example}

\numberwithin{equation}{thm}

\DeclareMathOperator*{\Wr}{\wr}

\DeclareMathOperator{\XOR}{\scriptstyle{\mathsf{XOR}}}
\DeclareMathOperator{\OR}{\scriptstyle{\mathsf {OR}}}
\DeclareMathOperator{\AND}{\scriptstyle{\mathsf {AND}}}

\newcommand{\Z}{\mathbb Z}

\renewcommand{\:}{\colon}
\renewcommand{\>}{\rightarrow}

\usepackage[backref,
pagebackref,%
bookmarks=true,%
colorlinks=true]%
{hyperref}
\begin{document}

\hyphenation{appli-cat-ions cryp-to-gra-phy com-ple-xi-ty com-po-si-ti-ons 
dis-tan-ce ad-di-ti-on in-effect-ive multi-pli-cat-ion con-junct-ion 
com-pos-it-ion funct-ions Mau-rer ge-ne-ra-li-z-ed equi-pro-b-ab-le 
coun-ter-de-pen-dent}
%
%
\def\huh{\hbox{\vrule width 2pt height 8pt depth 2pt}}
\def\eqnum#1{\eqno (#1)}
\def\cwdash{\relbar\joinrel}
\def\fnote#1{\footnote}

\title[Pseudorandom generators: an addendum] {Pseudorandom Number Generation by $p$-adic Ergodic
Transformations: An Addendum}
\author{Vladimir Anashin}

\address{Faculty of Information Security, 
Russian State University for the Humanities,\\
Kirovogradskaya Str., 25/2, Moscow 113534, Russia}

\email{anashin@rsuh.ru, vladimir@anashin.msk.su}

\begin{abstract}

The paper study counter-dependent pseudorandom number generators 
based on $m$-variate
($m>1$)
ergodic mappings of the space of $2$-adic integers $\Z_2$.
The sequence of internal states of these generators is defined by the recurrence
law 
$\mathbf x_{i+1}= H^B_i(\mathbf x_i)\bmod{2^n}$, whereas their output
sequence is
$\mathbf z_{i}=F^B_i(\mathbf x_i)\mod 2^n$; here
$\mathbf x_j, \mathbf z_j$ are $m$-dimensional vectors over $\Z_2$. It is shown how the
results obtained
for a univariate case could be extended to a multivariate case.
\end{abstract}

\maketitle

\section {Introduction}
\label{Sec:Intro}
In \cite{me:3} we considered counter-dependent generators that produce
recurrence sequences $\{u_i\in\Z/2^n\}$ of $n$-bit words according to the
following law:
$$u_{i}=F_i(w_i);\quad w_{i+1}\equiv f_i(w_i)\pmod{2^n},\quad (i=0,1,2,\ldots).$$
In the mentioned paper we restricted ourselves mainly to the case of univariate mappings $f_i$
and $F_i$. Trivially, each univariate mapping $\Z/2^{mn}\>\Z/2^{mn}$ 
of the resdue ring modulo $2^{mn}$ could be considered as a mapping 
$(\Z/2^n)^{(m)}\>(\Z/2^n)^{(m)}$ of a Cartesian power $(\Z/2^n)^{(m)}$
of the residue ring $\Z/2^n$, i.e., as an $m$-variate mapping. It turnes
out, however, that in some cases it is more effective to implement a univariate
mapping in its multivariate form to achieve better performance.
For instance, recently in \cite{KlSh:3}
there were constructed examples of multivariate $T$-functions with a single
cycle (i.e., of compatible ergodic functions, in our terminology, see \cite{me:3}),
which are very fast
(see theorem 6 of \cite{KlSh:3} and the text thereafter). 

Below we introduce some special way to derive multivariate compatible
ergodic functions from univariate ones (the mentioned mappings of 
\cite{KlSh:3} originate
this way); in fact, we merely represent univariate mappings in a multivariate
form. This immediately implies that {\slshape one could apply all the results
of \cite{me:3} to estimate important cryptographic characteristics of 
these multivariate mappings} (e.g.,
linear and $2$-adic spans, distribution of $k$-tuples), {\slshape as well as to construct
multivariate output functions that improve
periods of coordinate sequences} (see \cite{me:3} for definitions). Also,
exploiting this multivariate representation and 
using techniques
of wreath products of \cite{me:3}
we describe how to lift an arbitrary $m$-variate permutation with a single
cycle of $n$-bit
words to a permutation with a single cycle of $(n+K)$-bit words, and how
to construct counter-dependent generators based on these multivariate mappings.

\section{Multivariate ergodic mappings}
\label{Mult}
Consider a bijection
$B(x^0,\ldots,x^{m-1})=X$
of the $m$\textsuperscript{th} Cartesian  power $(\Z_2)^{(m)}$ of the space
$\Z_2$ of $2$-adic integers onto the space $\Z_2$ given by $\delta_k(X)\equiv\delta_{\ell}(x^r)\pmod
2$, where $r\in\{0,1,\ldots,m-1\}$ is the least non-negative residue of
$k\in\{0,1,2,\ldots\}$ modulo $m$, $k=\ell\cdot m+r$, $X\in\Z_2$,
$(x^0,\ldots,x^{m-1})\in(\Z_2)^{(m)}$, $\delta_j(u)$ is the 
$j$\textsuperscript{th} bit of a canonical $2$-adic representation of $u\in\Z_2$.
\footnote{Loosely speaking, we may
think of an element
of a Cartesian power $(\Z_2)^{(m)}$ as of a table of $m$ infinite binary
rows, to which we put into the correspondence an infinite binary string (that
is, an element of $\Z_2$) obtained
by reading succesively bits of each column, from top to bottom.} 
Consider a compatible mapping $H\:\Z_2\>\Z_2$ and a conjugate mapping
$$H^B(x^0,\ldots,x^{m-1})=(h^0(x^0,\ldots,x^{m-1}),\ldots,h^{m-1}(x^0,\ldots,x^{m-1}))$$
of $(\Z_2)^{(m)}$ to $(\Z_2)^{(m)}$;
that is, $H^B(x^0,\ldots,x^{m-1})=B^{-1}(H(B(x^0,\ldots,x^{m-1})))$.
Obviously, the conjugate mapping $H^B$ is compatible and ergodic
whenever the mapping $H$ is ergodic. For instance, let $H(X)=1+X$, then 
%
$$\delta_j(H(X))\equiv\delta_j(X)+\prod _{s=0}^{j-1}\delta_s(X)\pmod 2$$
(we assume the product over the empty set is $1$); then the conjugate $m$-variate
mapping is given by
\begin{multline*}
h^k(x^0,\ldots,x^{m-1})= x^k\oplus
\bigg(\bigg(\bigwedge_{s=0}^{k-1} x^s\bigg)
\wedge
\bigg(\bigwedge_{r=0}^{m-1}
((x^r+1)\oplus x^r)\bigg)\bigg)=\\
x^k\oplus
\bigg(\bigg(\bigwedge_{s=0}^{k-1} x^s\bigg)
\wedge
\bigg(\bigg(
\bigg(\bigwedge_{r=0}^{m-1}x^r\bigg)+1\bigg)\oplus 
\bigg(\bigwedge_{r=0}^{m-1}x^r\bigg)\bigg)\bigg)
\end{multline*}
for $k=0,1,2,\ldots,m-1$. Here, we recall, $\wedge$ (or $\AND$) is a 
bitwise conjunction\footnote{i.e.,
a bitwise multiplication modulo 2}, 
$\oplus$ (or $\XOR$) is a bitwise addition modulo $2$
(we assume that a bitwise conjunction $\wedge$ over the empty set is $-1$,
i.e., the string of all $1$'s). One
could construct various multivariate compatible ergodic mappings combining 
this representation with the ergodicity criterion. We recall the latter:  
\begin{thm}
\label{ergBool}
{\rm (see \cite[Theorem 3.13]{me:3})} 
A mapping $T\colon\mathbb Z_2\rightarrow\mathbb Z_2$ is
compatible and measure preserving\footnote{That
is, $T$ induces a permutation on $\Z/2^n$ for all $n=1,2,3,\ldots$} 
iff for each $i=0,1,\ldots$ the Boolean function 
$\tau^T_i=\delta_i(T)$
in Boolean variables $\chi_0,\ldots,\chi_{i}$ could be represented as Boolean
polynomial of the form
$$\tau^T_i(\chi_0,\ldots,\chi_i)=\chi_i+\varphi^T_i(\chi_0,\ldots,\chi_{i-1}),$$ 
where $\varphi^T_i$
is a Boolean polynomial. The mapping $T$ is compatible  and ergodic iff,
additionaly, the Boolean function
$\varphi^T_i$ is of odd weight, that is,
takes value $1$ exactly at the odd number of points 
$(\varepsilon_0,\dots,\varepsilon_{i-1})$, where
$\varepsilon_j\in\{0,1\}$ for $j=0,1,\ldots,i-1$. The latter takes place if and only
if $\varphi^T_0=1$, and the degree of the Boolean polynomial $\varphi^T_i$ for
$i\ge 1$ is exactly
$i$, that is, $\varphi^T_i$ contains a monomial
$\chi_0\cdots\chi_{i-1}$.
\end{thm}

For instance, theorem \ref{ergBool} implies that an arbitrary univariate
compatible and ergodic mapping $T$ gives rise to the $m$-variate compatible
and ergodic mapping $T^B=(t^0,\ldots,t^{m-1})$ of the form
$$t^k(x^0,\ldots,x^{m-1})= x^k\oplus
\bigg(\bigg(\bigwedge_{s=0}^{k-1} x^s\bigg)
\wedge
\bigg(\bigwedge_{r=0}^{m-1}
((x^r+1)\oplus x^r)\bigg)\bigg)
\oplus u^k(x^0,\ldots,x^{m-1}),$$
where 
\begin{equation}
\label{eq:EvenPar}
\sum_{(x^0,\ldots,x^{m-1})=(0,\ldots,0)}^{(2^r-1,\ldots,2^r-1)}
\delta_r(u^k(x^0,\ldots,x^{m-1}))\equiv 0\pmod 2
\end{equation}
for all $r=0,1,2,\ldots$.\footnote{such mappings $u^k$ are called {\it
even parameters}
in \cite{KlSh:3}}
With the use of these considerations we deduce from theorem 
\ref{ergBool} 
the following
\begin{prop}
\label{cor:WP:mult}
Let $f^j_s\:\Z_2\>\Z_2$ $(s\in\{0,1,\ldots, m-1\}$, $j=0,1,\ldots, m-1)$ be {\textup
(}univariate{\textup)}
ergodic functions, let
$g^j_s\:\Z_2\>\Z_2$ $(s\in\{0,1,\ldots, j-1\}$ , $j=1,2,\ldots, m-1)$ be 
{\textup(}univariate{\textup)}
measure-preserving functions.
Then the mapping
$$H^B(x^0,\ldots,x^{m-1})=(h^0(x^0,\ldots,x^{m-1}),\ldots,h^{m-1}(x^0,\ldots,x^{m-1}))$$
of $(\Z_2)^{(m)}$ onto $(\Z_2)^{(m)}$ such that
\begin{gather*}
h^0(x^0,\ldots,x^{m-1})=
x^0\oplus
\bigg(\bigwedge_{r=0}^{m-1}
(f^{0}_r(x^r)\oplus x^r)\bigg)
;\\
h^1(x^0,\ldots,x^{m-1})=
x^1\oplus\bigg(g^1_0(x^0)\wedge
\bigg(\bigwedge_{r=0}^{m-1}
(f^{1}_r(x^r)\oplus x^r)\bigg)\bigg)
;\\
\ldots \ldots \ldots\ldots\ldots\ldots\ldots\ldots\ldots
\ldots\ldots\ldots\ldots\ldots\ldots\ldots\ldots\ldots\ldots\ldots\ldots\ldots\\
h^{m-1}(x^0,\ldots,x^{m-1})=
x^{m-1}\oplus\bigg(\bigg(\bigwedge_{s=0}^{m-2} g^{m-1}_s(x^s)\bigg)
\wedge
\bigg(\bigwedge_{r=0}^{m-1}
(f^{m-1}_r(x^r)\oplus x^r)\bigg)\bigg)
\end{gather*}
is ergodic. That is, for all $n=1,2,\ldots$ the mapping $H$ 
induces modulo $2^n$ a permutation with a single
cycle; hence the length of this cycle is $2^{mn}$.
\end{prop}
\begin{proof}
It sufficies to demonstrate that the conjugate mapping 
$H\:\Z_2\>\Z_2$ is compatible
and ergodic.
Denote $\chi_k^r=\delta_k(x^r)$; we have to represent $\delta_t(h^s(x^0,\ldots,x^{m-1}))$
as a Boolean polynomial in Boolean variables $\chi_k^r$. For $c\in\{0,1,\ldots,
m-1\}$ let 
$$F^c=\bigwedge_{r=0}^{m-1}
(f^{c}_r(x^r)\oplus x^r);\qquad G^c= \bigwedge_{s=0}^{c-1} g^{c}_s(x^s), \quad
(c>0);\qquad
G^0=-1.$$
Now, since the functions $g_s^j$ and $f_s^j$ are compatible and,
respectively, measure preserving/ergodic, in view of \ref{ergBool} one
obtains the following representation of $\delta_k(g_s^j)$ and $\delta_k(f_s^j)$
as Boolean polynomials:
\begin{gather*} 
\delta_k(g_s^j(x^s))=\chi_k^s+\varphi_k^j(\chi_0^s,\ldots,\chi_{k-1}^s);\\
\delta_0(f_s^j(x^s))=\chi_0^s+1;\\
\delta_k(f_s^j(x^s))=\chi_k^s+\chi_0^s\cdots\chi_{k-1}^s+
\psi_k^j(\chi_0^s,\ldots,\chi_{k-1}^s)
\quad (k>0);
\end{gather*} 
where $\deg\psi_k^j(\chi_0^s,\ldots,\chi_{k-1}^s)<k$.
Further, since 
$$\delta_k(G^c\wedge F^c)\equiv\prod_{s=0}^{c-1}\delta_k(g_s^c(x^s))\cdot
\prod_{s=0}^{m-1}(\delta_k(f_s^c(x^s)+\delta_k(x^s))\pmod 2,$$
the above equations imply that
\begin{gather*}
\delta_0(G^0\wedge F^0)=1;\\
\delta_0(G^c\wedge F^c)=\chi_0^0\cdots\chi_0^{c-1}+\Phi_0^c, \quad (c>0);\\
\delta_k(G^0\wedge F^0)=\chi_0^0\cdots\chi_{k-1}^0\cdots
\chi_0^{m-1}\cdots\chi_{k-1}^{m-1}+
\Phi_k^0, \quad (k>0)
;\\ 
\delta_k(G^c\wedge F^c)=\chi_k^0\cdots\chi_k^{c-1}\cdot
\chi_0^0\cdots\chi_{k-1}^0\cdots
\chi_0^{m-1}\cdots\chi_{k-1}^{m-1}+\Phi_k^c,\quad (c>0, k>0). 
\end{gather*}
where $\Phi_k^c$ (respectively, $\Phi_k^0$ or $\Phi_0^c$) is a Boolean polynomial in
Boolean variables 
$$\chi_k^0,\dots,\chi_k^{c-1},
\chi_0^0,\dots,\chi_{k-1}^0,\dots,
\chi_0^{m-1},\dots,\chi_{k-1}^{m-1}$$ 
(respectively, in
$\chi_0^0,\dots,\chi_{k-1}^0,\dots,
\chi_0^{m-1},\dots,\chi_{k-1}^{m-1}$ or $\chi_0^0,\dots,\chi^{c-1}_0$), 
and $\deg\Phi_k^c<mk+c$.
Finally,
$
\delta_k(h^c(x^0,\ldots,x^{m-1}))=
\chi_k^c+\delta_k(G_k^c\wedge F_k^c), 
$
and the result follows in view of \ref{ergBool}.
\end{proof}
\begin{note}
\label{note:Mult:oplus}
Of course, the assertion of the proposition remains true for the mappings
$\hat h^s=h^s\oplus u^s$, $(s=0,1,\ldots,m-1)$, where $u^s$ is an arbitrary
mapping that satisfies \eqref{eq:EvenPar}, since these mappings $u^s$ add
summands of degree $<mk+s$ to each Boolean polynomial 
$\delta_k(h^s(x^0,\ldots,x^{m-1}))$, see the proof of \ref{cor:WP:mult}.
\end{note}
With this note we can deduce some consequences of proposition \ref{cor:WP:mult}.
\begin{cor}
\label{cor:Mult:KS}
{\rm \cite[Theorem 6 and Lemma 1]{KlSh:3}}
The $m$-variate mapping defined by
$$h^s(x^0,\ldots,x^{m-1})=x^s\oplus((h(x^0\wedge\cdots\wedge x^{m-1})\oplus
(x^0\wedge\cdots\wedge x^{m-1}))\wedge x^0\wedge\cdots\wedge x^{s-1}),$$
$s=0,1,\ldots,m-1$, is compatible and ergodic whenever $h$ is a univarite
compatible and ergodic function.
\end{cor}
\begin{proof} Just note that both
$\delta_k\big(\bigwedge_{t=0}^{m-1}(h(x^t)\oplus x^t)\big)$ and 
$\delta_k\big(h\big(\bigwedge_{t=0}^{m-1}x^t\big)\oplus
\big(\bigwedge_{t=0}^{m-1}x^t\big)\big)$ are Boolean polynomials of the
same degree $mk+s$.
\end{proof}
\begin{cor} 
\label{cor:Mult:plus}
For $m>1$ under conditions of \ref{cor:WP:mult} the following
$m$-variate mapping
$$
h^{t}(x^0,\ldots,x^{m-1})=
x^{t}+\bigg(\bigg(\bigwedge_{s=0}^{t-1} g^{t}_s(x^s)\bigg)
\wedge
\bigg(\bigwedge_{r=0}^{m-1}
(f^{t}_r(x^r)\oplus x^r)\bigg)\bigg),
$$
$t=0,1,\ldots, m-1$, is compatible and ergodic.
\end{cor}
\begin{proof} Integer addition $+$ adds carry from the $(mk+c)$\textsuperscript{th}
bit to $(m(k+1)+c)$\textsuperscript{th} bit of the coniugate mapping $H:\Z_2\>\Z_2$;
the carry is a Boolean polynomial in variables
$$\chi_k^c,\chi_k^0,\dots,\chi_k^{c-1},
\chi_0^0,\dots,\chi_{k-1}^0,\dots,
\chi_0^{m-1},\dots,\chi_{k-1}^{m-1},$$
hence, integer addition just adds a Boolean polynomial in $km+c+1$ variables  to the Boolean polynomial
$\delta_{k+1}(h^c(x^0,\ldots,x^{m-1})$ in $(k+1)m+c$ variables. So this
extra summand is of degree at most $km+c+1<(k+1)m+c$, see the proof of
proposition \ref{cor:WP:mult}.
\end{proof}
\begin{note} 
\label{note:Mult:plus}
Again, the corollary remains true for the mapping
$\hat h^s=h^s+u^s$, $(s=0,1,\ldots,m-1)$, where $u^s$ is an arbitrary
mapping that satisfies \eqref{eq:EvenPar}.
\end{note}

We recall 
that
according to \cite[Proposition 3.10]{me:3}, a compatible univariate function
$g\:\Z_2\>\Z_2$ (resp., $f\:\Z_2\>\Z_2$) preserves measure 
(resp., is ergodic)
iff
it could be represented as
$g(x)=d+x+2\cdot v(x)$  \textup {(}respectively as
$f(x)=1+x+2\cdot(v(x+1)-v(x))$\textup {)} for suitable $d\in\mathbb Z_2$ 
and compatible 
$v\colon\mathbb
Z_2\rightarrow \mathbb Z_2$. In other words, one can assume $v$ to be an arbitrary
(e.g., key-dependent)
composition of arithmetic operations (such as addition, multiplication,
subtraction, etc.) and bitwise logical operations (such as $\XOR$, $\AND$,
$\OR$, etc.); see \cite{me:3} for details. Thus, to obtain a cycle of length,
say, $2^{256}$ applying the above results, one could use $8$-variate mappings and work with $32$-bit
words, which are standard for most contemporary computers.

We note, however, that similarly to a univariate case, only senior bits
of output sequence achieve maximum period length: To
be more exact, if $x^j_i$ is the value of the $j$\textsuperscript{th} variable
at the $i$\textsuperscript{th} step, 
$(x^0_{i+1},\ldots,x^{m-1}_{i+1})=H^B(x^0_i,\ldots,x^{m-1}_i)$, then the
period length of the bit sequence $\{\delta_s(x^j_i)\:i=0,1,2,\ldots\}$ is 
$2^{ms+j+1}$,
for
$s\in\{0,1,\ldots\}$,
$j\in\{0,1,\ldots,m-1\}$. This could be improved by the use of multivariate
output functions in a manner of \cite[Proposition 4.13]{me:3}, namely: 
\begin{prop}
\label{pr:OutMult}
Let $H^B$ and$F^B$ be $m$-variate ergodic mappings that satisfy conditions
of proposition \ref{cor:WP:mult}, and let $\pi\:\Z/n\>\Z/n$ be an arbitrary
permutation of bits of $n$-bit word $z\in\Z/2^n$ such that $\delta_0(\pi(z))=\delta_{n-1}(z)$
{\rm (e.g., $\pi$ could be a bit order reversing permutation, or a $1$-bit
cyclic
shift towards senior bits)}. Consider a recurrence sequence $\mathcal Y=\{\mathbf
y_i\:i=0,1,2\ldots\}$ over $(\Z/2^n)^{(m)}$
defined by the laws
$$\mathbf x_{i+1}=H^B(\mathbf x_i)\bmod 2^n;\quad 
\mathbf y_i=F^B(\pi(x^{m-1}_i),x^0_i,\ldots,x^{m-2}_i)\bmod 2^n,$$
where $\mathbf x_j=(x_j^0,\ldots,x_j^{m-1}), 
\mathbf y_j=(y_j^0,\ldots,y_j^{m-1})\in(\Z/2^n)^{(m)}$. Then the
output sequence $\mathcal Y$ is purely periodic, its period
length is exactly $2^{nm}$, each element of $(\Z/2^n)^{(m)}$ occurs at the
period exactly once, and the period length of each coordinate sequence
$\delta_k(\mathcal Y^s)=\{\delta_k(y_i^s)\:i=0,1,2,\ldots\}$ 
is exactly $2^{nm}$. \footnote{Recall that according to \cite{me:3} the term 
``exactly" 
within this context means that the purely periodic binary sequence $\delta_k(\mathcal Y^s)$
has no periods of lengths less than $2^{nm}$.}
\end{prop}
\begin{proof} Immediately follows by application of \cite[Proposition 4.13]{me:3} 
to (univariate) conjugate mappings $H$ and $F$; we just note that Proposition
4.13 of \cite{me:3}, as it easily follows from its proof, 
holds for arbitrary permutation $\pi$ that satisfies conditions
of our proposition \ref{pr:OutMult}. 
\end{proof}
\begin{note}
As it follows from the proof of \cite[Proposition 4.13]{me:3}, to provide maximum
period length of all coordinate sequences of output sequence 
it is sufficient only to apply output function in such a way, that
the most significant bit of a state transition function substitutes for
the
least significant bit of argument of the output function. Thus, the proposition
\ref{pr:OutMult} remains true if one 
permutes variables $x^0,\ldots,x^{m-2}$ of the function $F^B$ in arbitrary
order, or permutes bits in these varibles, or apply arbitrary bijections
to these variables, etc.
\end{note}

It turnes out that with the use of techniques of wreath products of \cite{me:3}
it is possible to ``lift" an arbitrary permutation on $(\Z/2^n)^{(m)}$
with a single cycle
to $(\Z_2)^{(m)}$, 
i.e. to obtain
``really multivariate" permutations with a single cycle (in a somewhat 
``univariate manner", of course).
Recall the following theorem, which is a generalization of theorem \ref{ergBool}:
\begin{thm}
\label{pr:WP:even}
{\rm (\cite[4.3 and 4.4; or 4.10]{me:3})}
Let $T\colon\mathbb Z/2^M\rightarrow\mathbb Z/2^M$, $M\ge 1$, 
be an arbitrary permutation
with a single cycle, and
let the mappings $H_z(\cdot)\:\Z_2\>\Z_2$, $(z\in\Z/2^M)$ satisfy 
the following conditions: 
\begin{enumerate}
\item $\delta_i(H_z(x))\equiv \delta_i(x)+\rho_i(z;x)\pmod 2\ (i=0,1,2\ldots),$
where $\rho_i$ are Boolean functions in Boolean variables 
$\delta_r(z)$, $\delta_s(x)$
 $(r\in\{0,1,\ldots,M-1\}$, $s\in\{0,1,\ldots, i-1\})$, and $\rho_0(z;x)=\rho_0(z)$ does
not depend on $x$;
\item $\sum_{z=0}^{2^M-1}\rho_0(z)\equiv 1\pmod 2;$
\item $\sum_{z=0}^{2^M-1}\sum_{x=0}^{2^i-1}\rho_i(z;x)\equiv 1\pmod 2$,
$i=1,2,\ldots$
\end{enumerate} 
Then the mapping
$$W(x)=T(x\bmod{2^M})+2^M\cdot H_{x\bmod{2^M}}
\bigg(\Big\lfloor\frac{x}{2^M}\Big\rfloor\bigg)$$
is transitive modulo $2^k$ 
{\rm(that is, induces a permutation with a single cycle on the residue ring
$\Z/2^k$ modulo $2^k$)}
for all $k\ge M$.
\end{thm}
From here we deduce the following
\begin{prop}
\label{pr:Lift}
Let $T\:(\Z/2^n)^{(m)}\>(\Z/2^n)^{(m)}$ be an arbitrary {\rm (not necessarily
compatible)} $m$-variate mapping
with a single cycle, let $H^B\:(\Z_2)^{(m)}\>(\Z_2)^{(m)}$ be any $m$-variate
compatible ergodic mapping mentioned above {\rm (see \ref{cor:WP:mult}, 
\ref{note:Mult:oplus}, \ref{cor:Mult:KS}, \ref{cor:Mult:plus}, \ref{note:Mult:plus})}.
Then the $m$-variate mapping $W^B(\mathbf x)=T(\mathbf x\bmod 2^n)+
(H^B(\mathbf x)\wedge((-2^n)^{(m)}))$ of $(\Z_2)^{(m)}$ onto
$(\Z_2)^{(m)}$ 
induces
a permutation with a single cycle modulo $2^N$ for all $N\ge n$.
\end{prop}
Recall that a $2$-adic representation of $-2^n$ is an infinite binary string such
that first $n$ bits of it are $0$, and the rest are $1$. In other words,
$H^B(\mathbf x)\wedge((-2^n)^{(m)})$ takes
$\mathbf x=(x^0,\ldots,x^{m-1})$ to
$(h^0(\mathbf x)\wedge(-2^n),\ldots,h^{m-1}(\mathbf x)\wedge(-2^n))$, thus
sending to $0$ the first $n$ low order bits,
whereas
$\mathbf x\bmod 2^n=(x^0\bmod 2^n,\ldots,x^{m-1}\bmod 2^n)$ sends to $0$
all senior order bits, starting with the $n$\textsuperscript{th} bit 
(we start enumerate bits with $0$).
\begin{proof}[Proof of proposition \ref{pr:Lift}] The conjugate mapping
$W$ satisfies \ref{pr:WP:even} for $M=nm$ since all Boolean polynomials
$\delta_j(h^s(\mathbf x))$ are of odd weight, see the proof of \ref{cor:WP:mult}.
\end{proof} 

Concluding the section we just note that it is clear now how to construct counter-dependent
generators with the use of
the above multivariate ergodic mappings. Take, for instance, $M>1$ odd,
and take a finite sequence\footnote{which may be stored in memory, or may be generated
on the fly while implementing the corresponding generator} 
$$\{\mathbf c_j=(c_j^0,\ldots, c_j^{M-1})\:j=0,1,\ldots,M-1\}$$ 
of $m$-dimensional  vectors over $\Z/2^n$
such that the sequence of its first coordinates 
satisfy conditions of proposition 4.3 of \cite{me:3};
that is, $\sum_{j=0}^{M-1}c_j^0
\equiv 0\pmod 2$,
and the sequence $\{c_{j\bmod M}^0\bmod 2\:j=0,1,\ldots\}$ 
is purely periodic of period length
exactly $M$.  Then take arbitrary $m$-variate ergodic mappings $H_j^B$ and
$F_j^B$, $j=0,1,\ldots,M-1$  described above and consider recurrence
sequences defined by the laws
$$
\begin{array}{rcl}
\mathbf x_{i+1}&=&(\mathbf c_{i\bmod M}\oplus H^B_{i\bmod M}(\mathbf x_i))\bmod 2^n;\\
\mathbf y_i&=&(\mathbf F^B_{i\bmod M}(\pi(x^{m-1}_i),x^0_i,\ldots,x^{m-2}_i))\bmod 2^n,\\
\end{array}
$$
for $i=0,1,2,\ldots$, where $\pi$ satisfies conditions of \ref{pr:OutMult}.
Then the sequence of internal states $\{\mathbf x_i\}$ is purely periodic
of period length exactly $M\cdot 2^{nm}$, and each $m$-dimensional vector
over $\Z/2^n$ occurs at the period exactly $M$ times. The output sequence
$\mathcal Y=\{\mathbf y_i\}$ is also purely periodic of period length exactly $M\cdot 2^{nm}$, 
and each $m$-dimensional vector
over $\Z/2^n$ occurs at the period exactly $M$ times; moreover,
the period length of each coordinate sequence
$\delta_k(\mathcal Y^s)=\{\delta_k(y_i^s)\:i=0,1,2,\ldots\}$ 
is a multiple of $2^{nm}$, which
is not less than $2^{nm}$ and does not exceed $M\cdot 2^{nm}$. This conclusion
follows immediately by application of  \cite[Propositions 4.6 and 4.13]{me:3}
to conjugate mappings $H_j$ and $F_j$. The other counter-dependent generators (for  $M=2^k$
or arbitrary $M$) based on \cite[4.3, 4.4, 4.6 and 4.10]{me:3} could be constructed
by the analogy.

\section{Skew shifts and wreath products: a discussion}
\label{Skew}
The aim of this section is to make more transparent the core mapping underlying the 
constructions
introduced in \cite{me:3}, \cite{me:2}, \cite{me:1}, \cite{me:conf}, \cite{KlSh:1},
\cite{KlSh:2}, \cite{KlSh:3}, as well as \cite{me:ex} and even \cite{me:gr}.
This mapping is wreath product\footnote{this notion is more common for group
theory} of permutations; wreath product of permutations
is a special
case of  a skew product transformation\footnote{the latter notion is well known in dynamical systems
and ergodic theory}. We recall the most abstract definiton:
\begin{defn} Given two non-empty sets $X$, $Y$, a mapping $h\:X\>X$,
and a mapping
$H\:X\>Y^Y$, where $Y^Y$ \footnote{i.e., a Cartesian power of $Y$} is a set of 
all mappings of $Y$ into
$Y$.  Denote the action of $H$ as $(H(x))(y)=H_x(y)$
for $x\in X, y\in Y$. Then the {\it skew product transformation}  $H\Wr h $ is a mapping of 
a direct product
$X\times Y$ into itself such that $(H\Wr h)(x,y)=(h(x),H_x(y))$. 
\end{defn}
It is obvious that if $h$ is a bijection and all $H_x$, $x\in X$ are bijections,
then $H\wr h$ is a bijection. For instance, if $\star$ is a quasigroup
operation on $Y$ \footnote{that is, for all $a,b\in Y$ both equations $y\star a=b$ 
and $a\star y=b$ have unique
solutions in $y$}, $F\:X\>Y$ is an arbitrary mapping and $H_x(y)=y\star
F(x)$, then $H\wr h$ is bijective whenever $h$ is bijective. A classical
example in ergodic theory is skew shift on torus, which takes $(x,y)\in
(\mathbb T)^{(2)}$ to $(x\boxplus \gamma, y\boxplus \alpha(x))$, where
$(\mathbb T)^{(2)}$ is a $2$-dimensional torus (i.e., a Cartesian product
of a real interval $[0,1]$ onto itself); $\gamma,\alpha(x)\in[0,1]$, and
$\boxplus $ is addition modulo $1$ of reals of $[0,1]$. 

Another example
of imporance to cryptography is an $i$\textsuperscript{th} 
round permutation $R_i(k)$ of a Feistel network: This permutation
takes $(x,y)\in(\Z/2^n)^{(2)}$ to $(y\oplus f_i(k,x), x)$ (with 
$k$ being a key). Obviously, $R_i(k)$ is a composition of a skew shift $(x,y)\mapsto
(x,y\oplus f_i(k,x))$ and a permutation $\tau(x,y)=(y,x)$, which merely
changes positions of two concatented $n$-bit subwords in a $2n$-bit word. 
By the way, we used a construction  somewhat
resembling this permutation $R_i(k)$ in \ref{pr:OutMult}: In fact, 
from \ref{ergBool}
it is clear that a compatible mapping (or a $T$-function, in
terminology of \cite{KlSh:1}) of $\Z/2^N$ into $\Z/2^N$ 
is 
a composition
of
$N$ skew product transformations of $\Z/2$, and that a measure preserving
mapping (or invertible $T$-function) is a skew shift on $N$-dimensional
discrete torus  $(\Z/2)^{(N)}$. The skew products seems to become popular
in cryptography: Boaz Tsaban  noted that a construction
of a counter-dependent generator of \cite{ShTs} is just an ergodic-theoretic 
skew-product of a counter (or any automata) with the given automata. 
In particular, if the counter is replaced by any ergodic transformation, 
then the resulting cipher will be ergodic, \cite{Ts}. All these observations
lead to a suggestion that there are tight connections between ergodic theory
and cryptography. In fact, in this pper we use the notions of ergodicity and measure
preservation just because the corresponding mappings are ergodic or measure-preserving
in exact sence of ergodic theory. 

Of course, the most intriguing is a question, which naturally arises in this connection,
whether an ergodic theory could give something to prove (or to give strong
evidence of) cryptographic security
of a corresponding schemes. Might be, it is too early to put such a question
now, yet note that one of one-way candidates, namely, DES with a fixed
message, is a composition of skew shifts with a permutation $\tau$. Note
that in a corresponding construction \cite{LuRa} DES is assumed to be a family of
pseudorandom functions. In \cite{me:3} we conjectured that a mapping $F\:\Z/2^n\>\Z/2^k$
defined by $k$ randomly and independently choosen Boolean polynomials (with
polynomially restricted number of monomials) in
$n$ variables is a one-way function, and gave some evidence that
among the
generators we studied there may exist ones that are provably strong
against a known plaintext attack. A stronger assumption that $F$ is
a pseudorandom function\footnote{to be more exact, assuming that it is
possible to construct with these mappings $F$ a family of pseudorandom
functions; the corresponding construction, which is under study now, 
is based on skew shifts}(how plausible this asumption is?) may lead to
a proof that a corresponding generator is pseudorandom. For instance,
forming of output sequence $\{y_i\}$ (see \cite[Section 6]{me:3}
for notations) a sequence
$y_{0}, y_{0}\oplus y_1,\ldots,y_{m-2}\oplus y_{m-1},\ldots$
with probability $1-\epsilon$ one obtains that\footnote{we are using an opportunity
here to fix a misprint in \cite{me:3}}
$$y_{0}=F(z), y_{0}\oplus y_1=F(z+1),\ldots,y_{m-2}\oplus y_{m-1}=F(z+m-1),\ldots$$
Yet under assumptions that are made, this sequence, as well as the output
sequence must be pseudorandom.

More ``ergodic-theoretic common features" could be seen while analysing proofs 
of corresponding reslts. The mappings defined by compositions of arithmetic
and bitwise logical operations turnes out to be continuous on $\Z_2$, and
moreover, rather close to uniformly differentiable mappings, see \cite{me:1},
\cite{me:2}, \cite{me:3}, \cite{me:conf}. To study certain important
cryptographic properties of these mapping we approximate them (with respect
to a $2$-adic distance) by uniformly differentiable functions; we have
to calculate derivatives of these functions to check whether a given mapping
is a permutation, or whether it is equiprobable. On the other hand, to
study similar questions for other algebraic systems, e.g., discrete groups,
we have also to study derivatives, namely, Fox derivatives of mappings
of groups, see \cite{me:gr}, \cite{me:ex} for details.  Thus, we have to
use ``continuous" techniques to study ``discrete" problems. We could continue
such observations. At our view,
all this is more than a mere analogy between ergodic-theoretic and cryptographical
constructions.

\end{document}